\documentclass[aps,prb,twocolumn,superscriptaddress]{revtex4-1}

\usepackage{graphicx}
\usepackage{dcolumn}
\usepackage{bm}
\usepackage{amssymb}

\begin{document}



\title{Unconventional Magnetic Ground State in Yb$_2$Ti$_2$O$_7$}



\author{R.~M.~D'Ortenzio}
 \affiliation{Department of Physics and Astronomy, McMaster University, 1280 Main St. W., Hamilton, ON, Canada, L8S 4M1}
\author{H.~A.~Dabkowska}
 \affiliation{Brockhouse Institute for Materials Research, McMaster University, Hamilton, Ontario L8S 4M1, Canada}
\author{S.~R.~Dunsiger} 
 \affiliation{Physik-Department E21, Technische Universitat Munchen, D-85748 Garching, Germany}
\author{B.~D.~Gaulin}
 \affiliation{Department of Physics and Astronomy, McMaster University, 1280 Main St. W., Hamilton, ON, Canada, L8S 4M1}
 \affiliation{Brockhouse Institute for Materials Research, McMaster University, Hamilton, Ontario L8S 4M1, Canada}
 \affiliation{Canadian Institute for Advanced Research, Toronto, Ontario, Canada, M5G 1Z8}
\author{M.~J.~P.~Gingras}
 \affiliation{Canadian Institute for Advanced Research, Toronto, Ontario, Canada, M5G 1Z8}
 \affiliation{Department of Physics and Astronomy, University of Waterloo, Waterloo, ON N2L 3G1, Canada}
\affiliation{Perimeter Institute for Theoretical Physics, 31 Caroline North, Waterloo, Ontario, N2L 2Y5, Canada}
\author{T.~Goko}
 \affiliation{Department of Physics, Columbia University, New York, New York 10027, USA}
\author{J.~B.~Kycia}
\affiliation{Department of Physics and Astronomy, University of Waterloo, Waterloo, ON N2L 3G1, Canada}
\author{L.~Liu}
\affiliation{Department of Physics, Columbia University, New York, New York 10027, USA}
\author{T.~Medina}
\affiliation{Department of Physics and Astronomy, McMaster University, 1280 Main St. W., Hamilton, ON, Canada, L8S 4M1}
\author{T.~J.~Munsie}
 \affiliation{Department of Physics and Astronomy, McMaster University, 1280 Main St. W., Hamilton, ON, Canada, L8S 4M1}
\author{D.~Pomaranski}
 \affiliation{Department of Physics and Astronomy, University of Waterloo, Waterloo, ON N2L 3G1, Canada}
\author{K.~A.~Ross}
 \affiliation{Institute for Quantum Matter and Department of Physics and Astronomy, Johns Hopkins University, Baltimore, Maryland 21218, USA}
 \affiliation{NIST Center for Neutron Research, National Institute of Standards and Technology, Gaithersburg, Maryland 20899, USA}
\author{Y.~J.~Uemura}
 \affiliation{Department of Physics, Columbia University, New York, New York 10027, USA}
\author{T.~J.~Williams}
 \affiliation{Department of Physics and Astronomy, McMaster University, 1280 Main St. W., Hamilton, ON, Canada, L8S 4M1}
\author{G.~M.~Luke}
 \email[]{luke@mcmaster.ca}
 \affiliation{Department of Physics and Astronomy, McMaster University, 1280 Main St. W., Hamilton, ON, Canada, L8S 4M1}
 \affiliation{Brockhouse Institute for Materials Research, McMaster University, Hamilton, Ontario L8S 4M1, Canada}
 \affiliation{Canadian Institute for Advanced Research, Toronto, Ontario, Canada, M5G 1Z8}


\date{\today}

\begin{abstract}
We report low temperature specific heat and positive muon spin relaxation/rotation ($\mu$SR) measurements on both polycrystalline and single-crystal  
samples of the pyrochlore magnet Yb$_2$Ti$_2$O$_7$.
This material is believed to possess a spin Hamiltonian able to support a Quantum Spin Ice (QSI) ground state.
Yb$_2$Ti$_2$O$_7$  displays sample variation in its low temperature heat capacity and,
while our two  samples exhibit extremes of this variation, 
 our $\mu$SR measurements indicate a similar disordered low temperature state down to 16 mK in both.  We report little temperature dependence to the 
muon
spin relaxation and no evidence for ferromagnetic order, in contrast to reports  by Chang \emph{et al.} [Nat. Comm. {\bf 3}, 992 (2012)] and Yasui \emph{et al.} [J. Phys. Soc. Japan. {\bf 72}, 11 (2003)].  
Transverse field (TF) $\mu$SR  measurements show changes in the temperature dependence of the muon Knight shift 
that coincide with heat capacity anomalies which, incidentally, 
prove that the implanted muons are not diffusing in Yb$_2$Ti$_2$O$_7$. 
From these results, we are  led to propose that Yb$_2$Ti$_2$O$_7$ enters an unconventional ground state below $T_c\sim265$~mK.
As found for all the current leading experimental candidates for a quantum spin liquid  state, the
precise nature of the state below $T_c$ in Yb$_2$Ti$_2$O$_7$ 
remains unknown and, at this time, defined by what it is not as opposed to what it is:
lacking simple periodic long-range order or a frozen spin glass state.
\end{abstract}


\maketitle

\section{Introduction}
The R$_2$Ti$_2$O$_7$ rare-earth titanates 
(R is a trivalent magnetic rare earth ion and Ti$^{4+}$ is non-magnetic) 
offer outstanding opportunities for the study of geometric magnetic frustration.
The R sites within the cubic pyrochlore structure, with space group $Fd\overline{3}m$, 
form a three-dimensional network of corner sharing tetrahedra.
This structure has a high propensity towards frustrated magnetic interactions 
and allows for a wide range of exotic low temperature magnetic states and phenomenologies~\cite{Gardner2}.
For example, classical spin ice pyrochlore materials such as Dy$_2$Ti$_2$O$_7$ and Ho$_2$Ti$_2$O$_7$,
which have been extensively studied and are now relatively well understood~\cite{Gardner2,Gingras_Springer}, 
 have been proposed to exhibit monopole-like deconfined quasiparticle excitations~\cite{C_Castelnovo}.


A new and exciting research direction is that of {\em quantum spin ice}, 
a state in which quantum fluctuations  may elevate the classical spin liquid state \cite{Balents_Nature} of spin ice 
 to a full-blown quantum spin liquid (QSL) \cite{Hermele,Molavian,Onoda,Balents,Lee_Onoda_Balents,savary_2013}. 
A possible candidate for a quantum spin ice is Yb$_2$Ti$_2$O$_7$
\cite{Bramwell,Hodges3,Hodges,Gardner3,Hodges2,Reotier,Chang,
Yasui,K_Ross4,Thompson,K_Ross1,K_Ross3,K_Ross2,Applegate,Hayre}
which does not exhibit the gradual spin freezing and residual low-temperature entropy of a classical spin ice\cite{Applegate,Pomaranski}.
 The nature of Yb$_2$Ti$_2$O$_7$'s ground state is currently under much debate, with some  experiments
 reporting a ferromagnetic low temperature state~\cite{Chang,Yasui}, 
while others show a dynamic ground state with no long range order~\cite{Hodges,K_Ross3,K_Ross4,Gardner3} 
and which may be compatible with a QSL state.
At the same time, on the theoretical front,
the microscopic Hamiltonian for Yb$_2$Ti$_2$O$_7$ appears quite well described by a pseudospin-1/2 
quantum spin ice model~\cite{K_Ross1,Applegate,Hayre} which is predicted within conventional\cite{K_Ross1,Wong} as well as gauge\cite{Applegate} mean-field theories to possess a splayed ferromagnetic ground state similar to that recently reported for Yb$_2$Sn$_2$O$_7$\cite{Yaouanc_Sn} for the microscopic parameters
obtained from inelastic neutron scattering studies of Yb$_2$Ti$_2$O$_7$ ~\cite{K_Ross1}.



Previous zero field (ZF) $\mu$SR~\cite{Hodges, Hodges2, Reotier} 
performed on polycrystalline Yb$_2$Ti$_2$O$_7$ showed no evidence of 
long-range magnetic order below a sharp specific heat transition at $T_c \sim$240~mK. 
Rather, an Yb$^{3+}$  spin fluctuation rate drop of approximately 4 orders
 of magnitude was observed at $T_c$ ~\cite{Hodges, Hodges2}. M\"{o}ssbauer absorption spectroscopy measurements 
by the same group also indicated a first order magnetic transition, but these did not allow for a determination of  the ground state.

Different neutron scattering experiments have found
evidence for both an ordered ferromagnetic ground state~\cite{Chang, Yasui} and a dynamic magnetic ground state,
 with only short range ($3$D) spin correlations~\cite{K_Ross3, K_Ross4}. 
Rods of magnetic scattering along the $\langle 111 \rangle$ directions have been reported~\cite{Hodges2, K_Ross3, K_Ross4, Thompson} 
 characteristic of  short range two-dimensional ($2$D) correlations above 400 mK \cite{K_Ross3}.  
Refs.~[\onlinecite{K_Ross3}, \onlinecite{K_Ross4}] found  that below 400~mK, the rods begin to coalesce and build up intensity near the ${\bm Q} = (1,1,1)$ 
nuclear Bragg peak, interpreted~\cite{K_Ross3} as a cross-over to short range three-dimensional ($3$D) magnetic correlations, but clearly lacking long range order. 
Inelastic neutron scattering at 30~mK in these same studies \cite{K_Ross3,K_Ross4} 
shows only diffuse quasielastic scattering, and no evidence of sharp spin waves, again consistent with a dynamic disordered ground state. 
In contrast with Refs.~[\onlinecite{K_Ross3},\onlinecite{K_Ross4}],  Chang {\it et al.}'s
polarized neutron scattering results~\cite{Chang} show suppression of the $\langle 111 \rangle$ scattering rods evolving into 
 magnetic Bragg peaks in a first order manner below $T \sim$210~mK, but no concomitant 
presence of spin waves, as expected for such an ordered state,  has so far been reported.

Recent evidence suggests that, like other geometrically frustrated magnets such as Tb$_2$Ti$_2$O$_7$~\cite{Matsuhira}, 
the ground state of Yb$_2$Ti$_2$O$_7$ may be sensitive to small amounts of quenched disorder and non-stoichiometry at the $\sim$1$\%$ level, 
with several reports of sample variation of the large
heat capacity anomaly near 260~mK and below~\cite{K_Ross2,K_Ross3,Yaouanc}. 
Variation of the heat capacity is observed between samples, both poly- and single-crystalline, 
with the largest heat capacity anomaly appearing at the highest temperature
 taken as a figure of merit for the sample quality. To date, this anomaly is 
seen to be sharpest in polycrystalline samples that were prepared by the method outlined in Ref.~[\onlinecite{K_Ross2}].

Neutron diffraction studies~\cite{K_Ross2} have shown that a single-crystal of Yb$_2$Ti$_2$O$_7$ grown using the optical floating zone technique is weakly ``stuffed'', 
with stoichiometry Yb$_2$(Ti$_{2-x}$Yb$_x$)O$_{7-x/2}$ where $x\sim 0.046$, or 2.3$\%$ extra Yb$^{3+}$ ions 
reside on the nonmagnetic Ti$^{4+}$ sublattice. Meanwhile, polycrystalline samples prepared 
using the method in Ref.~[\onlinecite{K_Ross2}] are seemingly not stuffed  ($x \approx 0$).
These defects have been proposed~\cite{K_Ross2} to contribute to the sample variation in magnetic ground state and specific heat properties.

\begin{figure}[htb]
\includegraphics[scale=0.7]{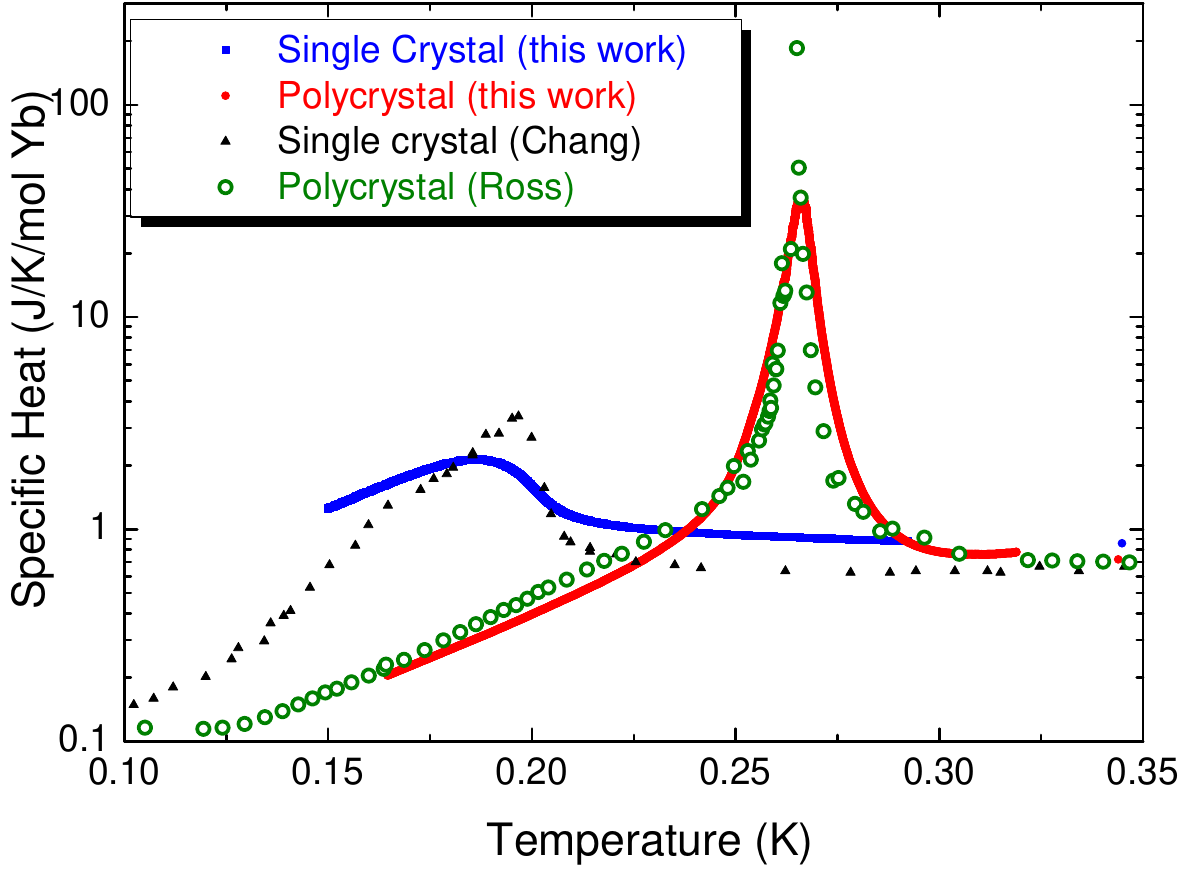}
\caption{\label{specific_heat} Low temperature specific heat measurements of Yb$_2$Ti$_2$O$_7$ on a semi$-$logarithmic scale. The red and blue curves represent samples used for this work. Our polycrystalline sample exhibits a sharp transition at $\sim$265 mK, and our single-crystal shows a broad peak at $\sim$185 mK. The black data is taken from Chang {\em et al.}'s~\cite{Chang} single-crystal, and the green is the sample~\cite{Reotier} used for Hodges 
{\em et al.}'s $\mu$SR measurements~\cite{Hodges}. 
The orange points are taken from Ross {\em et al.}~\cite{K_Ross3}.}
\end{figure}

In this paper, we study two samples of Yb$_2$Ti$_2$O$_7$ with very different low temperature heat capacities, shown in Fig.~\ref{specific_heat}. One is a polycrystalline sample with a sharp and large heat capacity anomaly at a relatively high $T_c$$\sim$265~mK. 
The other is a single-crystal sample which displays a broad anomaly at $T_c$$\sim$185~mK, with a peak amplitude 20 times smaller than that of the polycrystalline sample. For comparison, we also show the  specific heat measurements reported by Chang {\em et al.} for the single-crystal used 
in their neutron study~\cite{Chang}, Dalmas de R$\acute{\textup{{e}}}$otier {\em et al.}'s 
polycrystalline sample \cite{Reotier} 
used in a previous $\mu$SR study~\cite{Hodges} and Ross {\em et al.}'s polycrystal used in Ref.~[\onlinecite{K_Ross3}].  Based on our specific heat data, we would expect 
our single-crystal to have a relatively high degree of stuffing  as the specific heat peak is pushed down to $\sim$185~mK. 
Our polycrystalline sample shows a sharp peak at the highest temperature so far reported, 
from which we infer that it should be closer to balanced stoichiometry ($x\approx$0). Remarkably, the $\mu$SR measurements we report below
show little difference between the two samples at low temperatures, and both the single-crystal and our high quality polycrystalline sample are shown to remain disordered and dynamic 
down
to the lowest temperatures 
considered.



\section{Results}

We prepared polycrystalline pellets of Yb$_2$Ti$_2$O$_7$  by mixing stoichiometric quantities of Yb$_2$O$_3$ and TiO$_2$, isotropically pressing at 60 MPa then sintering at 1200~$^\circ$C for 24 hours. We then grew single-crystal Yb$_2$Ti$_2$O$_7$ from some of the polycrystalline material  using the optical floating zone crystal growth method in 2 ATM of O$_2$ as described in  Ref.~[\onlinecite{springer}].
SQUID measurements of the dc-magnetization above 2~K confirm Curie-Weiss behavior with overall weak ferromagnetic interactions, $\Theta_{\textup{cw}}\approx 0.4$ K, similar to values reported in  previous studies~\cite{Bramwell, Hodges3}.

We performed specific heat measurements in a $^3$He/$^4$He dilution refrigerator at the University of Waterloo, using a combination of the relaxation and sweep methods. Pieces of the single crystal and polycrystalline samples  used in the muon spin relaxation measurements described later were sectioned into masses 45.5 mg and 48.0 mg, respectively. The samples were thermally connected to a temperature controlled stage through a Manganin wire, with respective thermal conductivities of $6.25\times10^{-8}$~J/K/s and $5.61\times10^{-8}$~J/K/s at 0.48 K. The average step size for the relaxation method was $2-10$\% of the nominal temperature, with an equilibration time window at least 3 times the weak link relaxation time constant.

\subsection{Longitudinal Field $\mu$SR}

\begin{figure}[htb]
\includegraphics[scale=0.7]{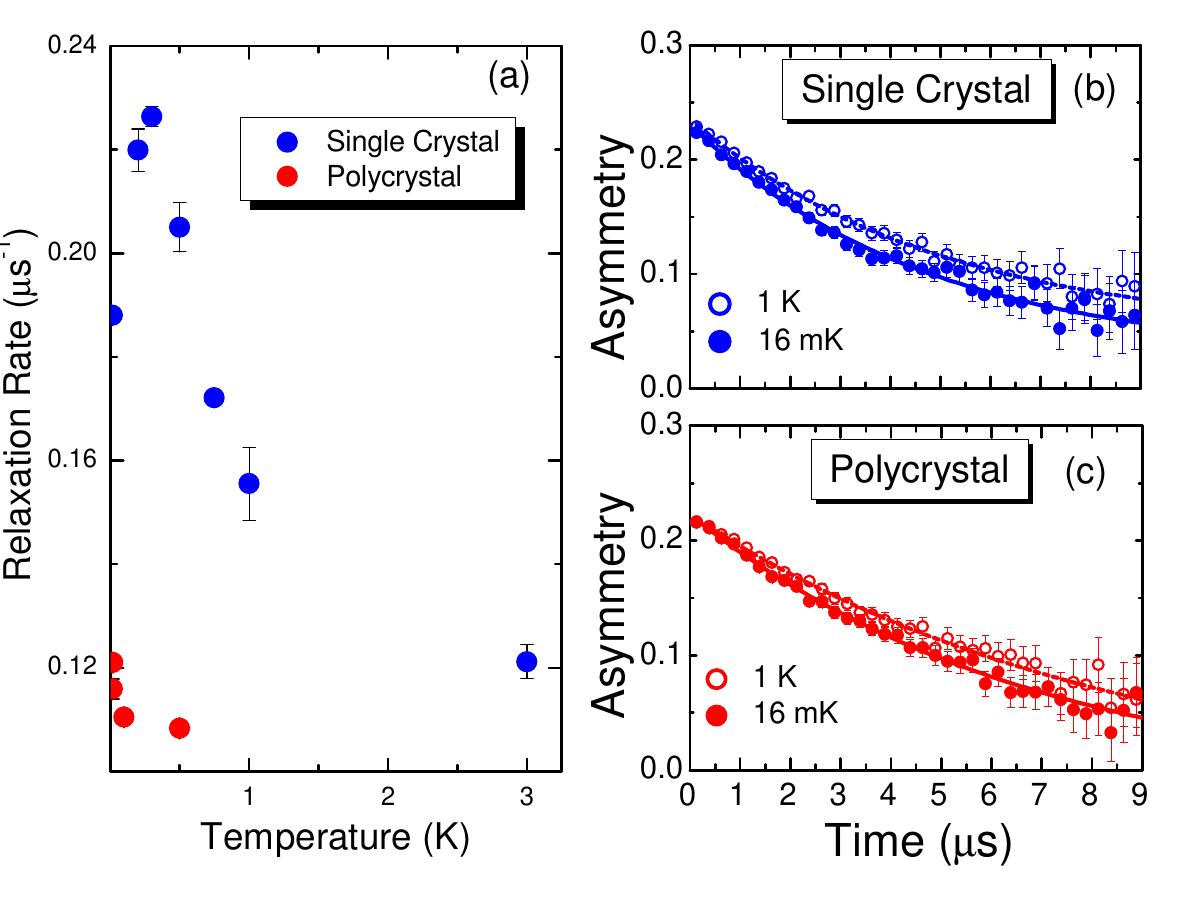}
\caption{\label{zf_data} A comparison of the weak LF (less than 0.5 mT) data for polycrystalline and single-crystal Yb$_2$Ti$_2$O$_7$.  Asymmetry spectra (shown in (b) and (c)) were fit with a single exponential decay with a globally fixed asymmetry for all temperatures. Plot (a) shows the temperature dependence of the relaxation rate. }
\end{figure}

Weak longitudinal field (LF) $\mu$SR spectra for $T=1$ K and 16 mK are shown in Fig.~\ref{zf_data} for both single-crystal and polycrystalline samples. 
Calibration spectra using transverse positron counters
show that  the  applied field is $H \leq$ 0.5 mT. 
We confirmed that the longitudinal relaxation rate did not vary with 
the strength of the weak applied field (up to at least $\sim$2 mT) indicating that the small field does not affect the dynamics of the Yb$^{3+}$ moments, although it does serve to decouple any static nuclear dipole relaxation from any stray muons landing in the cryostat tails or sample holder~\cite{Luke1}.

We find that  the entire muon polarization signal relaxes with a simple exponential decay 
characterized by 
a weak relaxation rate $\lambda(T)$ that is only slightly temperature dependent, as shown in Fig.~\ref{zf_data}. 
No oscillatory component or rapid relaxation that would be observed from a state with static moments on the $\mu$SR time scale is present down to 16 mK for either sample, noting that the $\mu$SR time scale is some six orders of magnitude longer than that relevant to the neutron scattering experiment\cite{Reotier1} finding ferromagnetic order \cite{Chang}.
We can preclude the possibility of a static local field lying solely along the initial muon polarization direction (a situation where no muon precession would be observed) in the polycrystal (where all crystal orientations are present) and in the single crystal (where the cubic symmetry would give 3 equivalent $\langle111\rangle$ directions, two of which would be perpendicular to the initial muon polarization).  
In marked contrast to the  results presented in Ref.~[\onlinecite{Hodges}], we observe no fast 
relaxing component of the signal at short time
for either of our samples at any temperature, including  below the  transition (polycrystal) or broad feature (single crystal) observed in the specific heat
measurements in Fig.~1.
This behavior allows us to conclude that the  Yb$^{3+}$ spins are in the fast fluctuating (narrowing) regime, 
with an absence  of  conventional long range magnetic order  or any  static internal magnetic fields such as would be present in a spin glass 
state
at low temperature.  The relaxation rate does increase somewhat with decreasing temperature, 
indicating a gradual slowing down of Yb$^{3+}$ spin fluctuations; however 
these
remain rapidly fluctuating down to 16~mK.
 
 \subsection{Transverse Field $\mu$SR}
\begin{figure}[htb]
\includegraphics[scale=0.75]{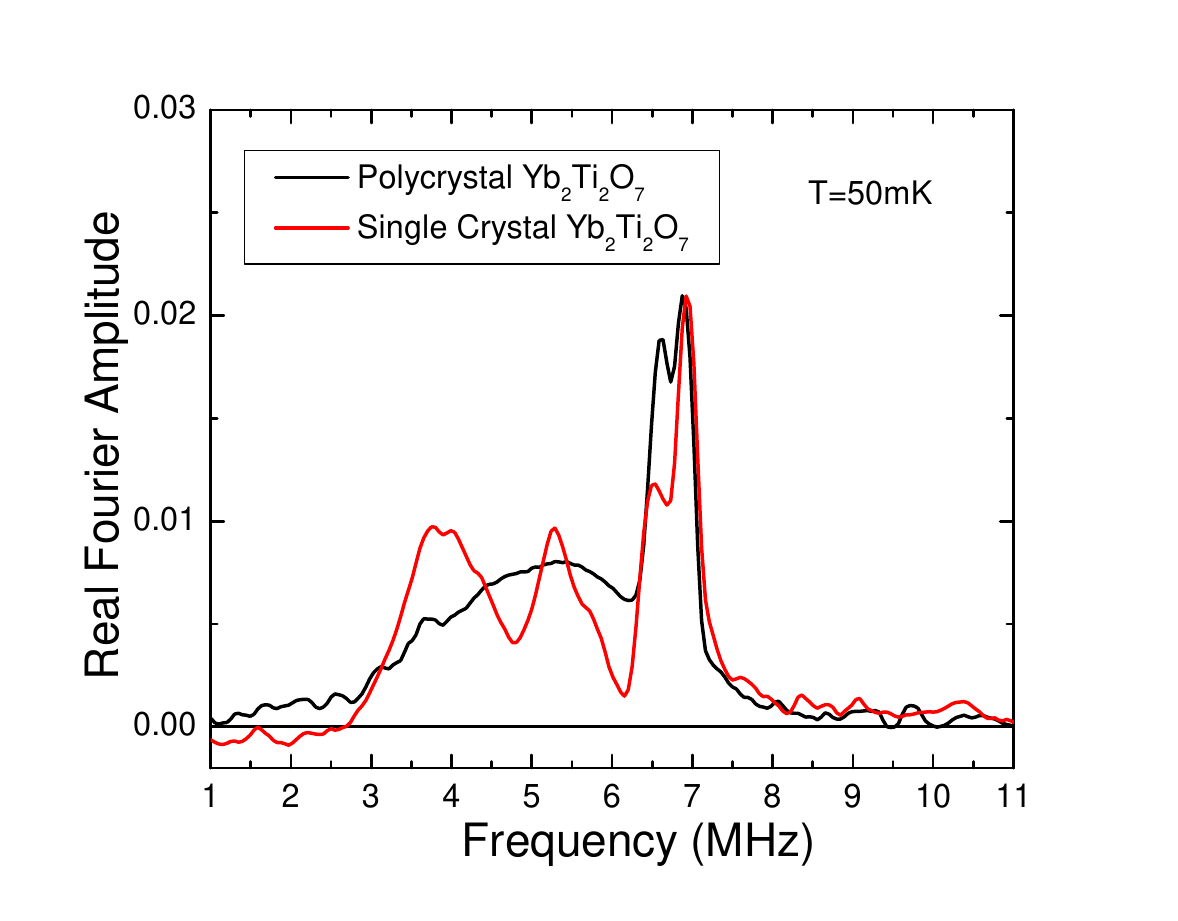}
\caption{\label{fft_tf}Fourier Transform of $\mu$SR asymmetry spectra in a transverse field of 50 mT at $T=50$ mK. The single-crystal Yb$_2$Ti$_2$O$_7$  shows 4 resolvable frequencies (red) and the polycrystalline Yb$_2$Ti$_2$O$_7$ shows three (black). This motivates $n=4$ for the single-crystal and $n=3$ in the polycrystalline Yb$_2$Ti$_2$O$_7$ for Eq. (1).  The sharp signal at approximately 7 MHz for both samples reflects the precession of the muons in the Ag cryostat tails.}
\end{figure}

\begin{figure}[htb]
\includegraphics[width=\columnwidth]{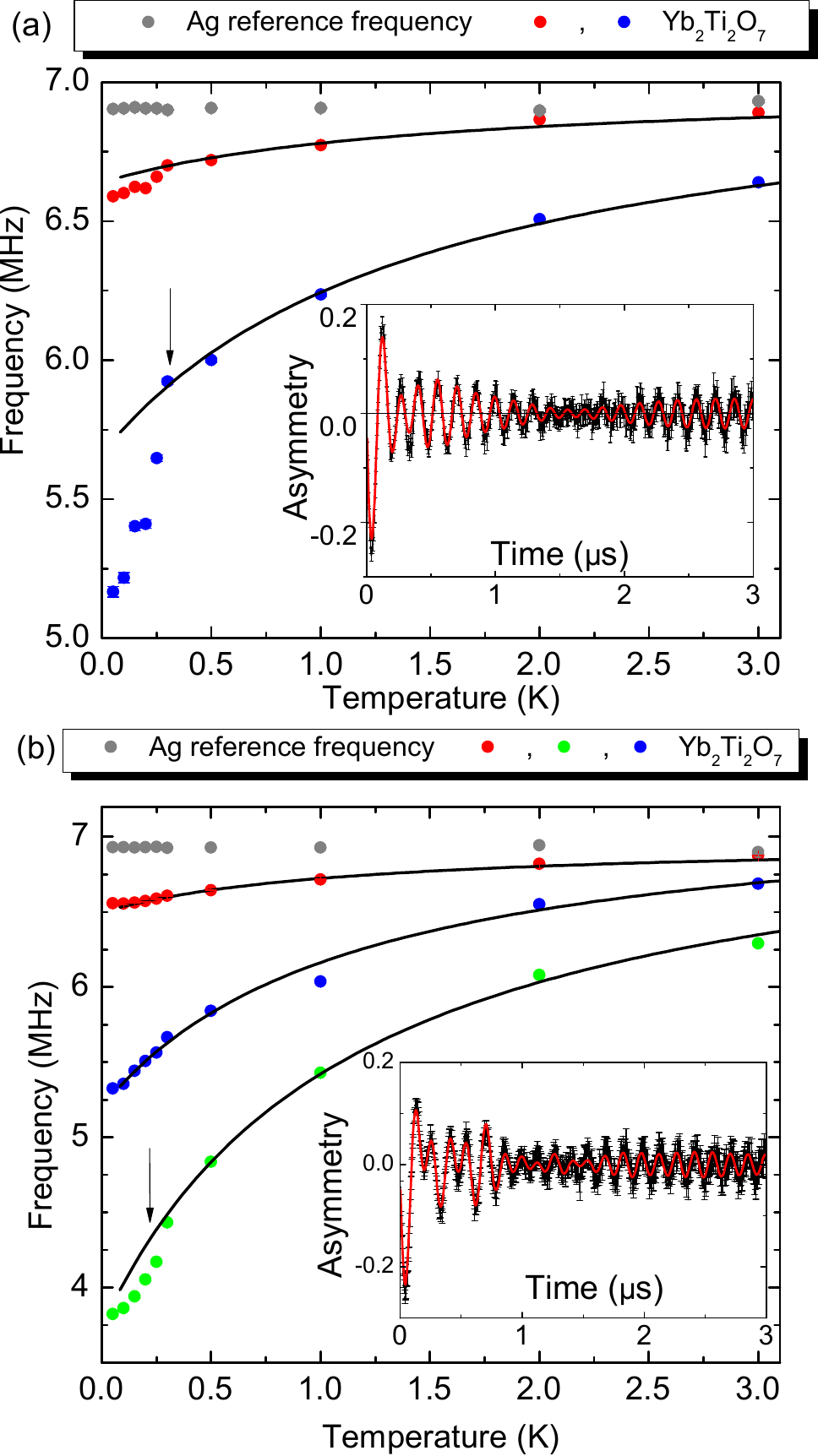}
\caption{\label{tf_both} The Yb$^{3+}$ spin susceptibility characterized by TF-$\mu$SR measurements of (a) polycrystalline and (b) single-crystal Yb$_2$Ti$_2$O$_7$. For the single-crystal Yb$_2$Ti$_2$O$_7$ the initial muon  polarization points perpendicular to the TF$=50$~mT field which was applied parallel to
 $[$111$]$. There are two distinct muon signals from the polycrystalline sample, and three from the single-crystal sample. The arrows show the approximate temperature of the onset of the transition observed in the specific heat. The solid black lines represent Curie-Weiss fits, as described in the text. The insets show asymmetry spectra (black) and fit to Eq.~1 (red).}
\end{figure}

To further characterize the low-temperature state of Yb$_2$Ti$_2$O$_7$,
 we performed transverse field (TF) $\mu$SR measurements to probe the local spin susceptibility, applying $H=50$ mT  perpendicular to  the initial muon spin polarization, which was rotated 90$^\circ$ relative to the LF measurements. 
For the single-crystal Yb$_2$Ti$_2$O$_7$, the field  was applied parallel to $[111]$. A field of $H=50$~mT along $[110]$ is small enough not to induce a magnetically ordered state  and likely also for the $[111]$ direction\cite{K_Ross3, K_Ross4,Discuss_Hayre}. Fig.~\ref{fft_tf} shows Fourier transforms of time spectra measured at $T=50$ mK, which exhibit a number of resolved lines for each sample.  The positive muon site in materials  is determined by electrostatic interactions, with  muons  in Yb$_2$Ti$_2$O$_7$ most likely to reside near O$^{2-}$ ions\cite{Luke1}.  There are two crystallographically inequivalent  O$^{2-}$ ions in the pyrochlore structure\cite{Gardner2}, both of which could provide  muon sites.  The application of an external magnetic field along a specific (in this case a $[111]$) axis polarizes the Yb$^{3+}$ moments, making several crystallographically equivalent sites magnetically inequivalent, which 
therefore increases
the multiplicity  of muon precession frequencies. 
The splitting of each line from the frequency corresponding to  the applied field reflects the local (anisotropic) susceptibility 
probed at that particular muon site.   
The polycrystalline sample exhibits fewer lines, though they are broader in frequency as a result of the powder-averaging of the anisotropic shifts of different crystallographic sites.  After examining the Fourier transforms we chose to fit the time spectra to Eq.~\ref{eq1} with $n=3$ (polycrystal) and $n=4$ (single-crystal), fixing the amplitudes of the individual components at all temperatures for each sample to values obtained from simultaneous global fits over a range of temperatures.   Typical fits are shown in the insets of Fig.~\ref{tf_both} (a) and (b) for the polycrystal and single-crystal samples, respectively.

\begin{equation}
\label{eq1}
A(t)=\sum_{i=1}^{n}A{_{i}}e^{-\sigma {_{i}^{2}}t^{2}}\cos(\omega{_{i}}t+\varphi_i)
\end{equation} 

We show  the results of this analysis for the frequencies of each component  in Fig.~\ref{tf_both}.  The highest frequency component (labeled in grey) for each sample exhibits no temperature dependence. We ascribe this signal to muons landing in the silver sample holder which provides a useful reference signal for determining the muon Knight shift.  
The frequencies originating from the samples  all show a  strong temperature dependence as shown in Fig~\ref{tf_both}.  We parametrized the Knight shifts of these signals referenced to the silver signal by fitting them to a Curie-Weiss temperature dependence in the temperature range 400~mK~$<T<3.5$~K. We used a common ``$\mu$SR Curie-Weiss temperature'', $T^\mu_{\textup{cw}}$, for each frequency signal from a given sample, but allowed $T^\mu_{\textup{cw}}$ to vary between samples.  The resulting frequencies obtained from 
these fits of the Knight shifts  are indicated by the black solid lines in Fig.~\ref{tf_both}; for {\it both samples} 
we find $T^\mu_{\textup{cw}}=-1.3\pm0.5$~K, which is considerably below the value ($\Theta_{\textup{cw}}$=+0.4 K) we obtain at higher temperatures (above 2~K) from dc-magnetization measurements.   For both samples 
we see a  large and rather abrupt deviation from
this parametrized Curie$-$Weiss behavior  below about $T\approx0.25$~K.  This temperature  corresponds to  the onset of the specific heat phase transition in the polycrystal and is near to that of the single crystal, both indicated by
the vertical arrows in Fig.~\ref{tf_both}. The deviation is most pronounced in the polycrystalline sample where the specific heat transition is also the most
pronounced.  The change in the temperature dependence of the Knight shift indicates that there is a marked change in either  the local spin susceptibility and/or the Yb$^{3+}$-muon hyperfine coupling,
onsetting around the thermodynamic phase transition in Yb$_2$Ti$_2$O$_7$ as identified by the specific peak in Fig.~2.

The similarities of our data between the polycrystalline and single-crystal samples 
is surprising considering their very different specific heat signatures indicating 
a wide range of stuffing~\cite{K_Ross2}. It is also surprising that we see no evidence 
for the dramatic first-order-like change in the spin fluctuation rate that was previously reported~\cite{Hodges}.
This difference is apparent in the raw data  of Figs.~\ref{zf_data}b and \ref{zf_data}c and 
 is evidence of a sample dependence between those studied by Hodges {\em et al.}\cite{Hodges,Hodges2}
and this work which must go  beyond the differences in stuffing between our polycrystal and single-crystal samples. 

Most importantly, our results suggest that,  from a local microscopic point of view, 
the magnetic ground state is to a large extent insensitive 
to some aspects of sample variation and that a sharp specific heat transition does not signify magnetic order.

Our weak LF measurements, shown in Fig.~\ref{zf_data} exhibit a weakly temperature dependent spin fluctuation rate down to the lowest temperature considered.
  Many frustrated pyrochlore systems exhibit  a substantial, largely temperature independent LF/ZF 
relaxation~\cite{Uemura, McClarty, Luke1,Quemerais} which, although often seen in frustrated or 
low-dimensional magnetic systems, have not yet been explained~\cite{McClarty}. 
A recently proposed explanation  for  these apparent persistent spin dynamics in terms 
of quantum diffusion of muons, and suggested for Dy$_2$Ti$_2$O$_7$ ~\cite{Quemerais}, 
is untenable here (and likely in that case as well as we would expect similar muon diffusion
behavior  in the two systems).  If muons were mobile in Yb$_2$Ti$_2$O$_7$, they would most likely  move 
between different magnetically inequivalent sites which would broaden all of 
the individual TF lines in Fig.~3 into one, which clearly does not occur.  A recent ZF-$\mu$SR study of Y$_2$Ti$_2$O$_7$ also concluded
that quantum diffusion was absent in that related material\cite{Rodriguez}.
 Furthermore, we can exclude a ``muon impurity'' effect in which the presence of the muon would
 locally destroy the magnetic behavior, since our TF-$\mu$SR measurements are clearly  
sensitive to the transition  which strongly affects our measured muon Knight shift (see Fig.~4).

\section{Discussion}
Yb$_2$Ti$_2$O$_7$ in the polycrystalline (and stoichiometric)
form exhibits a clear thermodynamic phase transition 
evidenced by specific heat measurements 
that is accompanied by the marked change in the muon Knight shift in Fig. 4a. 
On the other hand, the broad
anomaly observed in the single crystal heat capacity may be indicative 
of  a crossover, rather than a thermodynamic phase transition, 
and consistent with the smoother shift seen in Fig. 4b compared to Fig. 4a.
We did not perform a detailed search for
hysteresis at this transition which precludes us from making a definitive statement regarding its order.
 The specific heat anomaly in the polycrystal is extremely sharp which could be evidence for a first order transition, 
 but further measurements will be required to confirm this. 
This leaves the nature of the low temperature state unresolved, although we do know that 
it involves at least a change of  the local spin susceptibility and/or hyperfine coupling, rather than complete magnetic ordering.  
This behavior reminds that of the moderate heavy fermion system URu$_2$Si$_2$ which
 has a thermodynamic phase transition at 17.5~K to a so-called ``hidden-order'' state~\cite{Mydosh} 
without magnetic order and whose order parameter has remained unknown for more than 20 years of detailed investigation. 
In the case of URu$_2$Si$_2$, recent studies~\cite{Schmidt} indicate that the phase transition originates 
from a change in the hybridization between the localized electronic states associated with the uranium 
atoms and the conduction band, as it enters the heavy fermion state. 
As such a mechanism cannot occur in insulating Yb$_2$Ti$_2$O$_7$, we must seek another mechanism
for its phase transition to a state with an unidentified order parameter. 

When discussing ``hidden-order'', one often refers to the ordering 
of other multipoles than the magnetic dipole moment \cite{Multipolar_RMP}, 
with the electric quadrupolar (E2) being typically the most important one after the magnetic dipolar moment.
A transition that would involve the electrical quadrupole moments of Yb$^{3+}$ 
at the first order transition in Yb$_2$Ti$_2$O$_7$ would seem hard
to rationalize given that Yb$^{3+}$ is a Kramers ion with vanishing time even (e.g. quadrupolar) 
operator matrix elements projected in its low-energy crystal-field doublet 
\cite{Lee_Onoda_Balents} along with a lowest excited crystal field levels at 
an energy $\sim 600$ K very large compared to the energy scale of the relevant microscopic 
Yb$^{3+}$$-$Yb$^{3+}$ E2 interactions.
 That being said, numerous exotic ground states have been proposed~\cite{Balents,savary_2013} 
for effective spin-1/2 pyrochlore Kramers systems, such as 
Yb$_2$Ti$_2$O$_7$. These include a $U(1)$ quantum spin liquid (QSL) and a Coulomb ferromagnet
 in addition to conventional antiferromagnetic and ferromagnetic long-range order. 
Interestingly, the recent suggestion of a first order transition between a paramagnet 
and the $U(1)$ QSL\cite{savary_2013} could be consistent with what we observe in Yb$_2$Ti$_2$O$_7$.
Another possibility which could be considered is a valence bond solid (VBS). Of these states,
 the QSL and VBS would have no static magnetic moments, consistent with our results.
 However, it is unclear whether or how the known Hamiltonian for Yb$_2$Ti$_2$O$_7$ \cite{K_Ross1,Applegate,Hayre,Wong} 
could give rise to either of these states.  
Which of these phases or perhaps other more exotic states is realized in Yb$_2$Ti$_2$O$_7$ remains to be seen.


The search for materials with a QSL ground state is at the forefront of theoretical and experimental condensed matter research\cite{Balents_Nature}.
Notwithstanding the paucity of such systems, one of the fundamental conceptual challenges facing the field is that the 
identification of a QSL ground state has so far typically been ``what it is not'' (i.e. lacking static multipolar order) rather than what it is\cite{Balents_Nature,What_not}.
Moving beyond this negative assessment of a QSL is one of the central questions being investigated in the field.
On the experimental side,  one still has to remain cautiously satisfied in obtaining a compelling demonstration of circumstantial 
evidence for a QSL state. An  example is the recently reported gapless fractionalized magnetic (spinons) excitations 
in the intensively studied ZnCu$_3$(OD)$_6$C$_{12}$ Herbertsmithite Cu$^{2+}$ $S=1/2$ kagome antiferromagnet \cite{Herbert}.

In our work, we have brought Yb$_2$Ti$_2$O$_7$ into the not very crowded 
arena of compelling QSL candidates by showing, as
a first essential step for further progress, that it is
an effective $S=1/2$ three-dimensional frustrated spin system lacking static dipolar order above 16~mK.
We hope that our work encourages others further systematic experimental and theoretical studies of this most interesting compound.

\begin{acknowledgments}
%
Research at McMaster University and University of Waterloo is supported by NSERC while research at Columbia is supported by the US NSF PIRE (OISE-0968226) and DMR-1105961 projects.  We appreciate the technical support of Dr. B. Hitti and Dr. A. Dabkowski and the TRIUMF Centre for Molecular and Materials Science. This work is supported in part by the Canada Research Chair program (M.J.P.G., Tier 1) and by the Perimeter Institute for Theoretical Physics through Industry Canada and by the Province of Ontario through the Ministry of Economic Development \& Innovation. 
\end{acknowledgments}

\end{document}